\documentclass[authoryear]{elsarticle}
\biboptions{square}
\biboptions{comma}

\usepackage{graphicx}

\usepackage{amssymb}
\usepackage{color}

\journal{Physics Letters B}

\usepackage{booktabs}

\begin{document}

\begin{frontmatter}



\title{Influence of $N^*$-resonances on hyperon production in the channel $\mathit{pp}\,\rightarrow\,K^+\!\Lambda p$   at 2.95, 3.20 and 3.30 GeV/c beam momentum}
\vspace{-5mm}
\author{\hspace{20mm}The COSY-TOF Collaboration}
\author[h]{S. Abd El-Samad}
\author[h]{M. Abdel-Bary}
\author[b]{K. Brinkmann\fnref{foot1}}
\fntext[foot1]{present address: Universit\"at Bonn, Germany}
\author[f]{H. Clement}
\author[b]{J. Dietrich}
\author[f]{E. Dorochkevitch}
\author[e]{S. Dshemuchadse}
\author[f]{K. Ehrhardt}
\author[f]{A. Erhardt}
\author{W. Eyrich\fnref{c}\corref{cor1}}
\cortext[cor1]{Corresponding author: W. Eyrich, Physikalisches Institut IV,
  Friedrich-Alexander Universit\"at Erlangen-N\"urnberg, Erwin-Rommel-Str. 1,
  91058 Erlangen, wolfgang.eyrich@physik.uni-erlangen.de}
\author[g]{C. Fanara}
\author[d]{D. Filges}
\author[g]{A. Filippi}
\author[b]{H. Freiesleben}
\author[c]{M. Fritsch\fnref{foot2}}
\fntext[foot2]{present address: Universit\"at Mainz, Germany}
\author[d]{W. Gast}
\author[c]{J. Georgi}
\author[d]{A. Gillitzer}
\author[b]{J. Gottwald}
\author[c]{J. Hauffe}
\author[d]{D. Hesselbarth}
\author[d]{H. J\"ager}
\author[b]{B. Jakob}
\author[b]{R. J\"akel}
\author[b]{L. Karsch}
\author[d]{K. Kilian}
\author[a]{H. Koch}
\author[c]{M. Krapp}
\author[f]{J. Kre\ss}
\author[b]{E. Kuhlmann}
\author[c]{A. Lehmann}
\author[g]{S. Marcello}
\author[d]{S. Marwinski}
\author[a]{S. Mauro}
\author[c]{A. Metzger}
\author[c]{W. Meyer}
\author[e]{P. Michel}
\author[e]{K. M\"oller}
\author[c]{H. M\"ortel}
\author[d,i]{H. P. Morsch}
\author[e]{L. Naumann}
\author[d]{N. Paul}
\author[c]{L. Pinna}
\author[c]{C. Pizzolotto}
\author[a]{C. Plettner}
\author[b]{S. Reimann}
\author[a]{M. Richter}
\author[d]{J. Ritman}
\author[d]{E. Roderburg}
\author[c]{A. Schamlott}
\author[b]{P. Sch\"onmeier}
\author[b]{M. Schulte-Wissermann}
\author[c,d]{W. Schroeder}
\author[d]{T. Sefzick}
\author[a]{M. Steinke}
\author[c]{F. Stinzing}
\author[b]{G. Sun}
\author[c]{A. Teufel}
\author[b]{W. Ullrich}
\author[c]{J. W\"achter}
\author[f]{G. J. Wagner}
\author[c]{M. Wagner}
\author[c]{R. Wenzel}
\author[a]{A. Wilms}
\author[d]{P. Wintz}
\author[c]{S. Wirth}
\author[d]{P. W\"ustner}
\author[i]{P. Zupranski}

\address[a]{Ruhr-Universit\"at Bochum, Germany}
\address[b]{Technische Universit\"at Dresden, Germany}
\address[c]{Universit\"at Erlangen-N\"urnberg, Germany}
\address[d]{Forschungszentrum J\"ulich, Germany}
\address[e]{Forschungszentrum Dresden-Rossendorf, Germany}
\address[f]{Universit\"at T\"ubingen, Germany}
\address[g]{INFN Torino, Italy}
\address[h]{Atomic Energy Authority NRC Cairo, Egypt}
\address[i]{Andrzej Soltan Institute for Nuclear Studies Warsaw, Poland}

\begin{abstract}
Hyperon production in the threshold region was studied in the reaction $\mathit{pp}\,\rightarrow\,K^+\!\Lambda p$ using the time-of-flight spectrometer COSY-TOF. Exclusive data, covering the full phase-space, were taken at the three different beam momenta of $p_\mathit{beam}=2.95$, $3.20$ and $3.30$\,GeV/c, corresponding to excess energies of $\varepsilon=204$, $285$ and $316$\,MeV, respectively.
Total cross-sections were deduced for the three beam momenta to be $23.9\pm0.8\,\pm2.0\,{\mu}b$, $28.4\pm1.3\,\pm2.2\,{\mu}b$ and $35.0\pm1.3\,\pm3.0\,{\mu}b$. Differential observables including Dalitz plots were obtained. The analysis of the Dalitz plots reveals a strong influence of the $N(1650)$-resonance at $p_\mathit{beam}=2.95$\,GeV/c, whereas for the higher momenta an increasing relative contribution of the $N(1710)$- and/or of the $N(1720)$-resonance was observed.
In addition, the $p\Lambda$-final-state interaction turned out to have a significant
influence on the Dalitz plot distribution.

\end{abstract}

\begin{keyword}
Associated strangeness production \sep Total cross-section \sep Dalitz plot

\PACS 14.20.Gk \sep 14.20.Jn \sep 14.40.Aq
\end{keyword}

\end{frontmatter}


\section{Introduction}

In order to obtain a complete and consistent picture of the structure and dynamics
of hadrons, precise measurements at medium and low energies are of fundamental
interest. In this context strangeness production in elementary reactions plays
an important role since the strange quark is not one of the constituent
quarks of the nucleon. To find out the relevant degrees of freedom of these
reactions in the near-threshold region,
an experimental program was started at the time-of-flight spectrometer
COSY-TOF \cite{COSY-TOF98,COSY-TOF06}.
\par
To describe the strangeness production in proton proton interaction various models based on meson exchange mechanism, including resonance effects, were published \cite{Laget91,Tsushima94,Sibirtsev95,Tsushima97,Faeldt97,Sibirtsev98,Shyam99,Sibirtsev00}.
In addition, an effective quark model has been used to calculate observables \cite{Kleefeld96}.
The principal distinction between the meson exchange models concerns the contribution of the strange and non-strange mesons, the consideration of the role of $N^*$-resonances and of the nucleon-hyperon ($\mathit{NY}$) and meson-hyperon ($\mathit{KY}$) interactions. However, theoretical stu\-dies do not yield a conclusive picture at present. Some stu\-dies are based on one boson (pion or kaon)
exchange amplitudes where kaon exchange was found to be the dominant process \cite{Laget91,Ferrari60}.
In other calculations the total cross-section data at somewhat higher
energies are well reproduced by pion exchange \cite{Yao62,Wu89}. In the resonance model approach
\cite{Tsushima94,Sibirtsev95,Tsushima97,Faeldt97,Sibirtsev98}, the
$K\Lambda$-production proceeds via the excitation of the $N(1650)S_{11}$-resonance
\cite{Tsushima97} or a combination of the $N(1650)S_{11}$-, $N(1710)P_{11}$- and
$N(1720)P_{13}$-resonances. In the calculation of \cite{Shyam99}, the $N(1650)S_{11}$- and $N(1710)P_{11}$-resonances 
dominate. A necessary condition to obtain conclusive results on the $K\Lambda$ production mechanism is the measurement of exclusive data covering the full phase-space.
\par
An essential part of the program at COSY-TOF aims at the production of $\Lambda$-,
$\Sigma^0$- and $\Sigma^+$-hyperons in proton-proton collisions and in a
further stage in proton-neutron reactions by use of a deuterium target. 
The COSY-TOF experiment at the external beam is designed to cover the momentum range from threshold up to the limit of the external COSY beam at about $3.4$\,GeV/c. The concept of the experiment lies in the complete geometrical
reconstruction of the events including the delayed decay of the
hyperons. Since the full phase-space of the primary reaction
products is covered by the detector and the events are kinematically complete, the total cross-section and all differential distributions can be extracted without extrapolations. 
\par
In a first run data of the reaction channel $\mathit{pp}\,\rightarrow\,K^+\!\Lambda p$ were taken at two beam momenta of $p_\mathit{beam}\,=\,2.50$ and $2.75$\,GeV/c. Total and differential cross-sections as well as the $\Lambda$-polarization and Dalitz plots were presented \cite{COSY-TOF98}. 
In a second publication \cite{COSY-TOF06} we reported on the results of a further $\Lambda-$production run at
three different beam momenta of $2.59$, $2.68$ and $2.85$\,GeV/c, corresponding to excess energies 
of $\varepsilon\,=\,85$, $115$ and $171$\,MeV, respectively.

In the paper presented here we report on the results of a $\Lambda-$production run covering the upper momentum region of COSY using three beam momenta of $2.95$, $3.20$ and $3.30$\,GeV/c, corresponding to excess energies of $\varepsilon\,=\,204$, $285$ and $316$\,MeV, respectively. 
Using an upgraded detector, significantly larger data samples have been collected. This allows more detailed analyses in particular of the Dalitz plots, which are the main topic of this paper.

\section{Experiment and Analysis}

\begin{figure}[ht]
\includegraphics[scale=0.6]{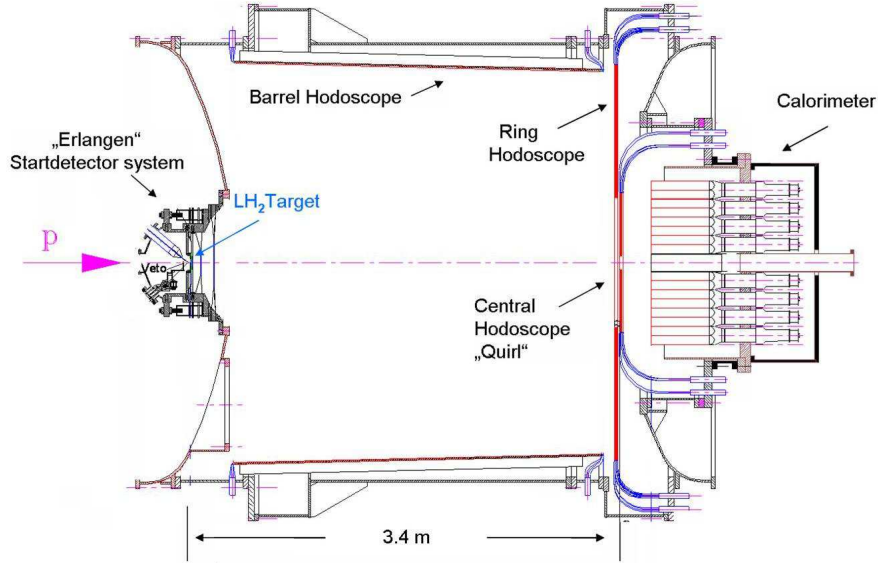}

\caption{\label{tof} \footnotesize{Schematic view of the COSY-TOF detector.}}
\end{figure}

\begin{figure}[ht]
\hspace{2,5cm}
\includegraphics[scale=0.4]{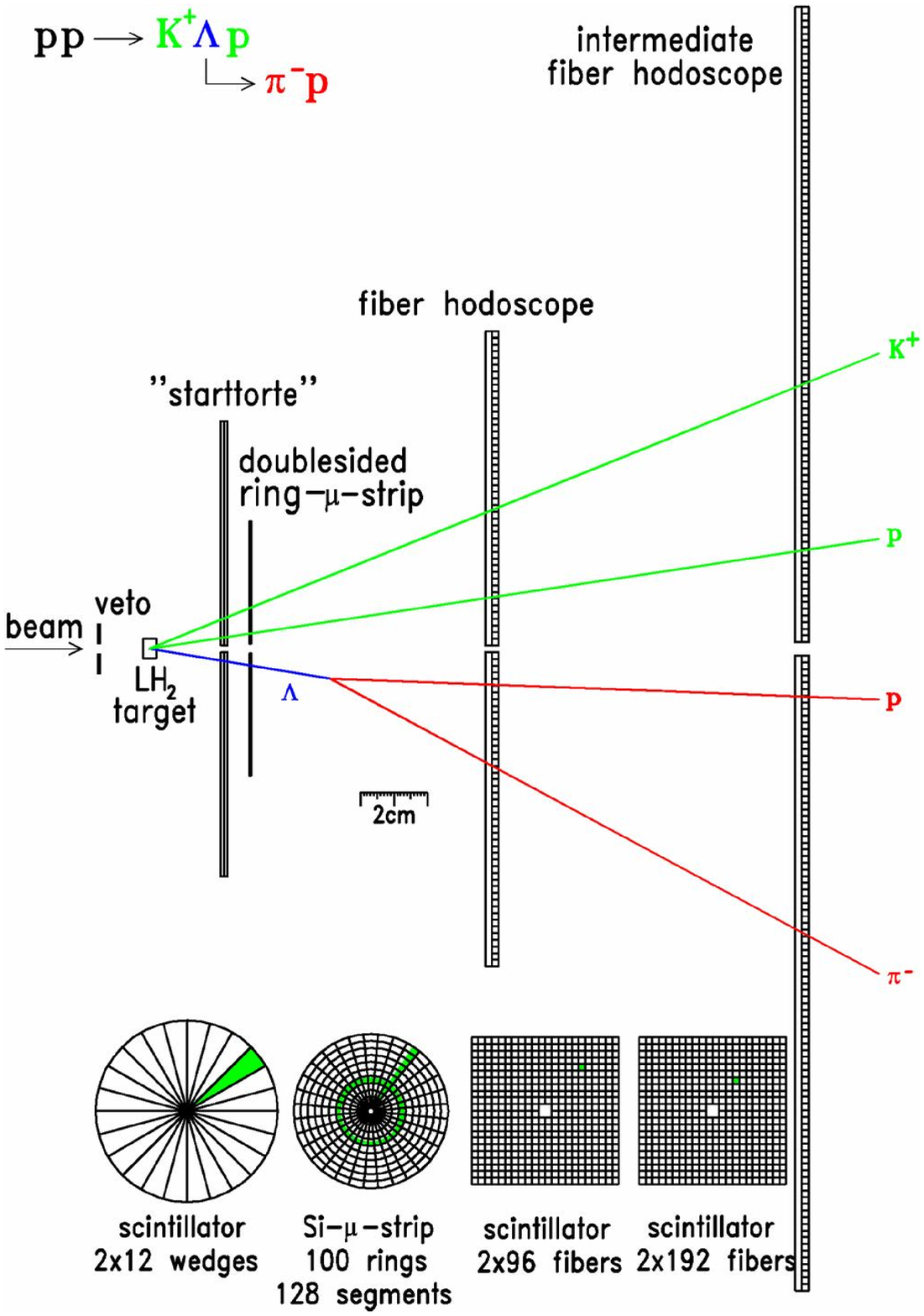}
\caption{\label{startcounter} \footnotesize{Schematic view of the start detector region with the track
pattern of a  $pp \rightarrow K^+\Lambda p$ event.}}
\end{figure}

The experiment was performed at the cooler synchrotron COSY with the COSY-TOF
spectrometer \cite{cosytofhomepage}, which is shown in Fig.\,\ref{tof}.
The reaction $pp \rightarrow K^+\Lambda p$ is induced by focusing the proton
beam on a spot of about 1 mm in diameter on a liquid hydrogen target (length 4
mm, diameter 6 mm). A schematic view of the start detector system is shown in Fig.\,\ref{startcounter} together with the track pattern of an event of the studied reaction. The first component, which is a thin scintillation counter with a diameter of 15 cm, consists of two segmented layers of 1~mm thickness.
Each layer is subdivided into 12 wedge shaped scintillators.  This
detector, called ``starttorte``, which is mounted 2 cm downstream of the target, provides the multiplicity of the charged particles close to the target as well as  the start signal for the time-of-flight measurement. It is
followed by a double-sided silicon  microstrip detector with a diameter of 6.2\,cm, the front and rear sides of which  consist of 100 rings and  128
sectors, respectively. This gives a precise determination of the track point close to the target with uncertainties given by the pixel size of $\Delta r=0.28$ mm and $\Delta \varphi = 2.8^o$. At a distance of 10 cm and 19 cm two scintillating fiber hodoscopes are installed to obtain two further track points. These detectors are essential to identify the tracks from the delayed decay of $\Lambda$ hyperons into a p\,$\pi^-$ pair. The first hodoscope consists of two crossed planes of 20~cm\,$\times$\,20~cm, built of square fibers (2~mm\,$\times$\,2~mm). The second hodoscope with a size of 40~cm\,$\times$\,40~cm consists of two crossed layers, built of the same type of fibers. 

\unitlength1cm

The outer detector system consists of the barrel hodoscope and an endcap with the central hodoscope and the ring hodoscope at a distance of about 3.4 m to the target (see Fig.\,\ref{tof}). The barrel hodoscope, which was not used in the former $\Lambda$-production runs\cite{COSY-TOF98, COSY-TOF06}, has a diameter of about 3 m and is built by 96 scintillating bars with a length of 2.8 m.
The outer detector provides the stop signal for the time-of-flight measurement as well as additional track points.
\unitlength1cm
The detector acceptance covers nearly the full phase-space of the primary particles ($\Lambda$-hyperon, $K^+$-meson, and proton) apart from holes with a diameter of about 2 mm for the start counter and the first hodoscope and about 4mm for the second hodoscope. These holes correspond to less than 3$\%$ of the phase-space. Kaons emitted at polar angles larger than the acceptance limit do not lead to a loss of information because of the forward-backward symmetry of the $pp$ entrance channel.
The trigger condition is based on the increase in charged-particle
multiplicity between the start scintillator and the outer detector system,
which results from the $\Lambda \rightarrow p\pi^-$ decay (see
Fig.\,\ref{startcounter}). Due to its high granularity the TOF spectrometer supplies good position information allowing precise track reconstruction. The track of the $\Lambda$-hyperon is reconstructed from the production vertex in the target, deduced using the primary tracks ($K^+$
and $p$), and the vertex of its decay into two charged
particles ($p$ and $\pi^-$).
\par
The high granularity of all detector components, covering nearly the full phase-space for the reaction of interest, allows the reconstruction of the events with sufficient precision using only the geometrical information
from the hit patterns in the various components. After the reconstruction of
the $\Lambda$-hyperon the momentum of each of the three primary particles is 
calculated from the measured directions using momentum conservation together 
with the beam momentum, which is known with a precision of about $\Delta p /p = 0.1 \%$ \cite{cosyref}. Assigning the primary charged tracks as $K^+$ and $p$, energy conservation allows the reconstruction of the mass of the neutral particle marked by the delayed decay. To improve the selection of the events and reduce combinatorial background, the measured energy loss of the ejectiles 
in the "starttorte" which gives a velocity resolution of about 8 $\%$ ($\sigma$) in the relevant range, was used in addition to the geometrical information of each primary track. Thus the possible wrong assignment of $K^+$ and $p$ tracks could be reduced to less than 4\%. The analysis was especially optimised to get a nearly background-free event sample. To control the analysis steps and deduce the reconstruction efficiency, extensive studies with simulated events were performed. The Monte Carlo simulation is based on Geant3.14 \cite{Geant} and the output of the digitalisation was fed into the same reconstruction algorithm used for the data. The overall reconstruction efficiency of events of the type $pp\to pK^+\Lambda$, $\Lambda\to p\pi^-$  was deduced to be between $3.6\%$ and $5.8\%$ for the different beam momenta (see table 1), 
nearly constant over the whole phase-space. The systematic error of the determination of the reconstruction efficiency  including the determination of the inefficiency of the detector system is estimated to be 5$\%$, based upon a comparison of three independant analyses \cite{cosytof2007}.
\unitlength1cm
\begin{figure}[ht]
\includegraphics[width=\textwidth]{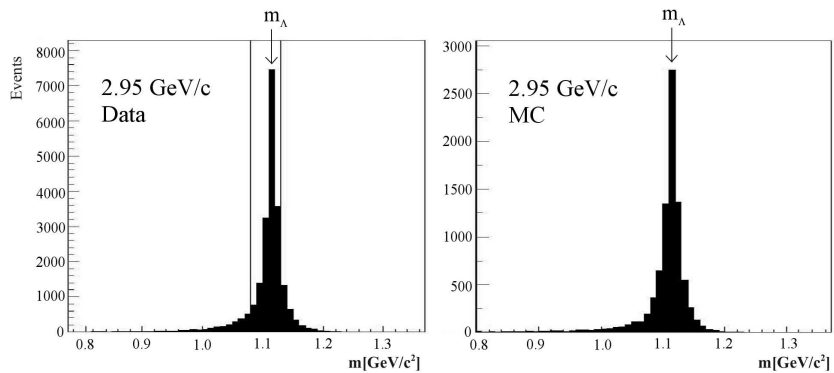}
\caption {\label{mmass} Left part: Missing mass distribution of $pp\to pK^+X$
at $p_{\mathit{beam}}\,=\,2.95$\,GeV/c. The selected range used for the further analysis is delimited by the vertical lines. Right part: Corresponding Monte Carlo simulation.}
\end{figure} 
\par
For the three beam momenta clean event samples were extracted. To determine the background, the $pK^+$ missing mass, which was extracted with a resolution of $16$\,MeV$/c^2$ (FWHM), was compared to Monte Carlo simulations. As an example in Fig.\,\ref{mmass} (left part) the missing mass spectrum at $p_\mathit{beam}\,=\,2.95$\,GeV/c is shown together with the cuts used to select the final event sample. 
In the right part the corresponding distribution of an analysed Monte Carlo sample is shown. From the comparison of both spectra a background contribution of about $6\%$ was deduced. The main part of the background consists of events of the type
$\mathit{pp}\,\rightarrow\,K^+\Sigma^0\,p$. From dedicated Monte Carlo
studies of the reaction $\mathit{pp}\,\rightarrow\,K^+\Sigma^0\,p$ the
contribution of these events in the final $\Lambda$-sample was determined to be
less than $4\%$ \cite{Schroeder}. As a last step, a kinematic fit with two overconstraints  was applied.
\par
The time integrated luminosity was calculated using the simultaneously measured elastic proton-proton scattering data which were compared with results from
\cite{EDDA}. The systematic error of the time integrated luminosities is 4$\%$ and corresponds to the errors in the cross-sections of the elastic scattering. In table 1 the number of events for the three measured beam momenta are given 
together with the corresponding reconstruction efficiencies and time integrated luminosities.
\par \noindent
\renewcommand{\arraystretch}{1.1}
\setlength{\tabcolsep}{6pt}
\begin{table}[h]
\begin{center}
\begin{tabular}{ccccc}
\toprule
$p_{\mathit{beam}}$ (GeV/c) & $\varepsilon$ (MeV) & $N_{\mathit{events}}$ & Efficiency ($\%$) & time integrated \\
 &  &  &  & Luminosity ($nb^{-1}$)\\

\midrule
2.95 & 204 & 15372 & 5.4 & 18.6\\
3.20& 285 & 5791 & 4.4 & 7.3\\
3.30& 316 & 6263 & 3.6 & 7.8\\
\bottomrule
\end{tabular}
\end{center}
\caption{Number of events for the measured beam momenta together with the corresponding reconstruction efficiencies and time integrated luminosities.}
\end{table}
\renewcommand{\arraystretch}{1.0}
\setlength{\tabcolsep}{6pt}
\par \noindent

\section{Results and Discussion}
\subsection{Total Cross-Sections}
The total cross-sections for the reaction $\mathit{pp}\,\rightarrow\,K^+\!\Lambda p$ are deduced from the selected data samples using the reconstruction efficiency from Monte Carlo analyses, including all relevant details of the detector and the luminosity as mentioned.

\begin{figure}[ht!]
  \includegraphics[width=\textwidth]{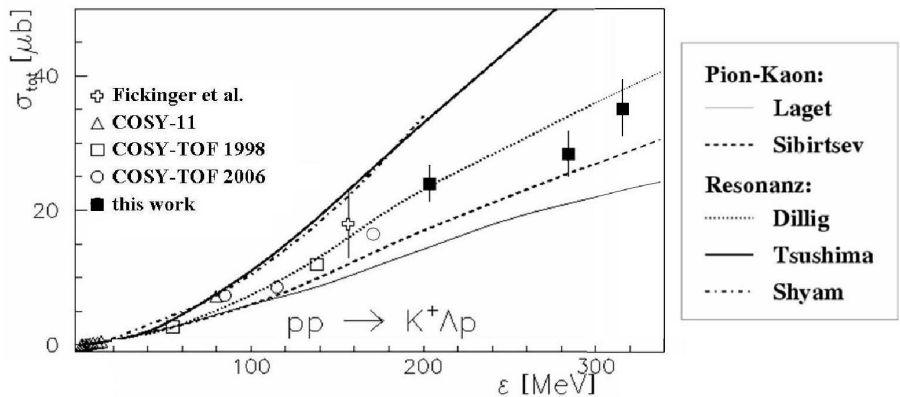}
  \caption{ \label{s_tot_modell} Total cross-sections of the channel $\mathit{pp}\,\rightarrow\,K^+{\!\Lambda}p$. All recent data up to $\varepsilon\,<\,340$\,MeV are shown \cite{COSY-TOF98,COSY-TOF06,COSY-1198,COSY-1199,Fickinger62} in comparison with model calculations: pion-kaon exchange model of Laget \cite{Laget91} and Sibirtsev \cite{Sibirtsev00}, resonance model of Dillig \cite{Kleefeld96}, Tsushima as given in \cite{Sibirtsev00} and Shyam \cite{Shyam99}. The error bars of the data points of this work are including statistical and systematical uncertainties.}  
\end{figure} 

The systematic error of the total cross-section is composed of the errors of the luminosity (4$\%$), the analysis (5$\%$), and the reconstruction efficiency determination (5$\%$). Total cross-sections and their statistical and systematic
errors were found to be $\sigma_\mathit{tot}=23.9\,\pm\,0.8\,\pm\,2.0\,{\mu}b$, $28.4\,\pm\,1.3\,\pm\,2.2\,{\mu}b$ and $35.0\,\pm\,1.3\,\pm\,3.0\,{\mu}b$ for excess energies $\varepsilon=204$, $285$ and $316$\,MeV, respectively. These data are shown by the solid squares in
Fig.\,\ref{s_tot_modell} as a function of the excess energy together with the
results of the experiment COSY-11 \cite{COSY-1198,COSY-1199} very close to
threshold, the previous data of COSY-TOF \cite{COSY-TOF98,COSY-TOF06} and 
an older data point from BNL \cite{Fickinger62}.
\par
In addition, the results of calculations obtained
from meson exchange models using different combinations of pion and
kaon exchange with and without resonance contributions are
plotted. The calculated cross-sections deviate among
each other and most of them do not reproduce the data in a satisfactory way. The best
agreement is achieved for the non-relativistic resonance model of Dillig \cite{Kleefeld96}
which does not apply any final-state interaction (FSI). The results of the
calculations of the pion-kaon exchange models (Laget \cite{Laget91} without and 
Sibirtsev \cite{Sibirtsev00} with $p\Lambda$-final-state interaction) more or less 
underestimate the data, while the results of the relativistic resonance models 
of Tsushima as given in
\cite{Sibirtsev00} and Shyam \cite{Shyam99} overestimate the data. 
But practically all of the models have adjustable parameters allowing then
 to describe the total cross section data. From that it is obvious that a comparison of the 
calculated and measured total cross-sections alone will not allow the extraction of the most relevant
 degrees of freedom for the investigated reactions in a unique way. To achieve a deeper insight into the reaction mechanism and into the properties of the hadrons involved, further experimental
 information, especially from differential observables, is needed. In this context, Dalitz plots are
 very promising representations of the differential data. 

\subsection{Dalitz Plot}
For reactions with more than two particles in the final-state, Dalitz plots are
a powerful tool to extract information about details of the reaction
mechanism. Whereas a pure phase-space distribution leads to a homogeneously populated
Dalitz plot, in particular contributing resonances should lead to significant
deviations. In the simplest case they appear as bands and accordingly as an enhancement in the projection on the squared mass of the respective two-body subsystem. Also final-state
interactions should show up as a characteristic enhancement in the Dalitz plot
and the low mass end of the corresponding projection of the related two-body
system. Due to interference effects it is not possible to extract the resonance and final state interaction strength from the projections alone. In general the Dalitz plot itself has to be analysed.
As the experiment covers practically the full phase-space, the Dalitz plot can be extracted in a model independent way. As discussed above, the background contamination is sufficiently small in the data presented here, therefore no background subtraction is needed.

\begin{figure}[ht!]
  \includegraphics[width=\textwidth]{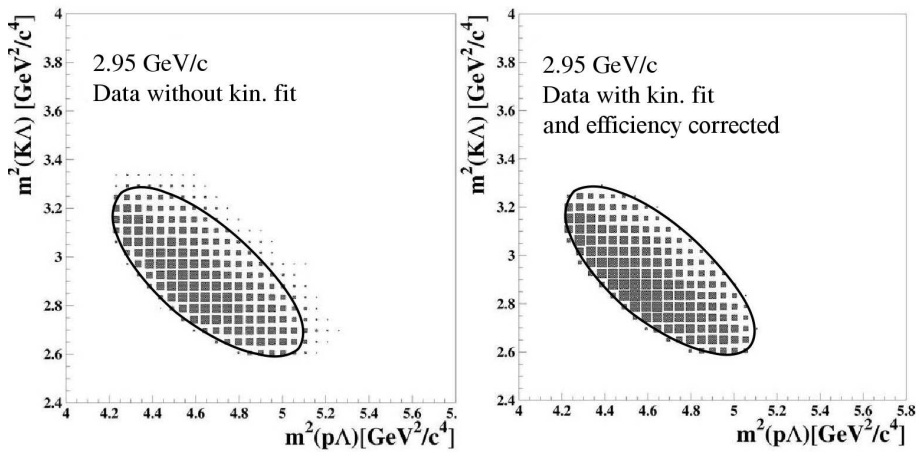}
  \caption{\label{dalitzplots} Dalitz plot at $p_{\mathit{beam}}\,=\,2.95$\,GeV/c, before (left) and after (right) efficiency correction and kinematic fit. Solid curves mark the kinematical limit.} 
\end{figure}

As an example, the Dalitz plot of the $pK^+\Lambda$ event sample at $p_{\mathit{beam}}\,=\,2.95$\,GeV/c is shown in the left part of Fig.\,\ref{dalitzplots}. As expected, the whole phase-space (indicated by the solid curve) is covered. The right plot corresponds 
to the same sample after applying the efficiency correction and a kinematic
fit. The resolution of the invariant mass in the
various subsystems is about $8$\,MeV$/c^2$ (FWHM). The comparison of these two figures show that the efficiency correction and the kinematic fit do not introduce significant artefacts.\par
The plots show strong deviations from a homogeneous
phase-space like distribution. A strongly enhanced area is visible in the left-lower
part. Monte Carlo simulations show that higher partial
waves with strength coefficients extracted from the experimental angular
distributions \cite{Schroeder}, do not influence the Dalitz plot distributions within the statistical errors of the given samples and therefore cannot be responsible 
for the observed strong deviation from a phase-space distribution. 
From our previous investigations \cite{COSY-TOF98, COSY-TOF06} and theoretical work, it is known that most of the observed anisotropy has its origin in the influence of $N^*$-resonances and
the $p\Lambda$-final-state interaction.

\subsection{Dalitz Plots Analysis}

To obtain more information on the various contributions, the data were compared
with the ISOBAR model prepared by Sibirtsev \cite{Sibirtsev} which
in turn is based on a concept outlined in \cite{SibHaid} and which was first applied to analyse the COSY-TOF data at $2.85\,$GeV/c \cite{COSY-TOF06}. In this approach which takes into account different $N^*$-resonances and the $p\Lambda$-final-state interaction the Dalitz plot distribution 
 $d{^2}\sigma/dm_{K\Lambda}^2 dm_{p\Lambda}^2$ is given as (see also \cite{PDG02} Eq. 38.21):
\[
\frac{d{^2}\sigma}{dm_{K\Lambda}^2 dm_{p\Lambda}^2}\,=\,
\mathit{fl\cdot ps}\,\cdot\,\left| \left(\sum_R\left(C_R \cdot A_R 
\right)\,+\,C_N \cdot A_N\right)\,\cdot\,\left(
1\,+\,C_{\mathit{FSI}} \cdot A_{\mathit{FSI}} \right)\,\right|^2 (1)
\]
Here $m_{K\Lambda}$ and $m_{p\Lambda}$ are the invariant masses of the $K\Lambda$-
and the $p\Lambda$-subsystem. The quantity $\mathit{fl}$ gives the normalization to 
the total cross-section, $p\mathit{s}$ represents the phase-space \cite{Byck}. The third factor gives 
the deviation from an uniformly populated Dalitz plot. $A_R$ are the relativistic Breit-Wigner-amplitudes of the three considered $N^*$-resonances. The corresponding values of mass and width which were chosen for the calculation are the following: $M=1650\,\textrm{MeV/c}^2$, $\Gamma=150\,\textrm{MeV/c}^2$; $M=1710\,\textrm{MeV/c}^2$, $\Gamma=100\,\textrm{MeV/c}^2$; $M=1720\,\textrm{MeV/c}^2$, $\Gamma=150\,\textrm{MeV/c}^2$. The sum of the weight factors $C_R$ and $C_N$ are normalized to 100. The amplitude of the non resonant background $A_{\mathit{N}} \equiv 1$ corresponds to a uniform phase-space. $A_{\mathit{FSI}}$ denotes the amplitude of the $p\Lambda$-final-state interaction as given by the J\"ulich $\mathit{YN}$-model \cite{Reu94}. The strength of the individual resonances $C_R$, of the non-resonant contribution $C_N$ (which includes the kaon exchange) and of the
$p\Lambda$-final-state interaction $C_{\mathit{FSI}}$ can be adjusted independently. 

\renewcommand{\arraystretch}{1.1}
\setlength{\tabcolsep}{3pt}
\begin{table}[ht!]
  \begin{center}
  \begin{tabular}{l|cccc}
  \toprule
  $p_{beam}$ & $C_{\mathit{R}}(1650)$ & $C_{\mathit{R}} (1710+1720)$ & $C_N$(phase-space) & $C_{FSI}(p\Lambda$)\\
  \midrule
  2.85 GeV/c (*) & $80\pm 10$ & $ 10\pm6$ & $10\pm8$ & $0.31\pm0.08$ \\
  2.95 GeV/c & $70\pm 8$ & $25\pm6$ & $5\pm6$ & $0.21\pm0.05$ \\
  3.20 GeV/c & $38\pm 8$ & $41\pm7$ & $21\pm6$ & $0.30\pm0.05$ \\
  3.30 GeV/c & $40\pm 9$ & $54\pm7$ & $6\pm6$ & $0.30\pm0.05$ \\
  \bottomrule
  \end{tabular}
  \caption{Values of $C_{\mathit{R/N/FSI}}$ obtained from the model adjustment. The given errors correspond to a 3 $\sigma$ interval. (*)The result of $2.85\,$GeV/c is taken from \cite{COSY-TOF06}.}
  \end{center}
\end{table}
\renewcommand{\arraystretch}{1.0}
\setlength{\tabcolsep}{6pt}
\begin{figure}[ht!]
  \includegraphics[scale=0.53]{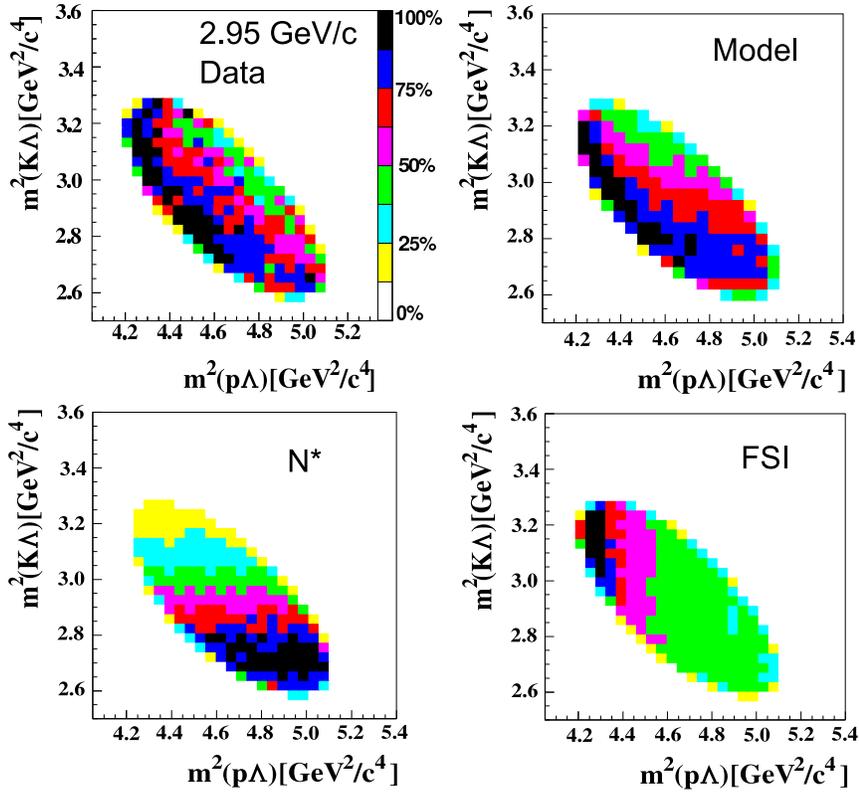}
  \caption{\label{mod_sib}Dalitz plot of the data at $p_{\mathit{beam}}\,=\,2.95$\,GeV/c (upper left), in comparison with the adjusted model of Sibirtsev (upper right), model calculation only with the resonance part without FSI (lower left), and only with $p\Lambda$-final-state interaction without resonances (lower right). In all plots, the scale from light to dark indicates the increasing yield in a linear scale.}
\end{figure} 
As already discussed in \cite{COSY-TOF06}, a satisfactory description of the data can only be achieved by taking into account the resonances and the final-state 
interaction together with a phase-space contribution at the amplitude level as given in equation (1). A comparison between the Dalitz plot of the data and the adjusted model (Fig.\,\ref{mod_sib} upper row) shows the best agreement with a dominant contribution of the $N(1650)$-resonance. From the small values of $C_N$ (see table 2) it has to be concluded, that the fraction of uniformly distributed events according to phase-space, including non-resonant processes especially kaon
exchange, is rather low. Moreover, the inclusion of the $p\Lambda$-final-state interaction to the calculation turned out to be absolutely necessary. 
For a reasonable description of the data a value of about $0.2$ to $0.3$ has to be taken for $C_{\mathit{FSI}}$. If only the resonances are taken into account, the characteristic horizontal band structure appears with a maximum at the position of the dominating $N(1650)$-resonance (Fig.\,\ref{mod_sib} lower left). A calculation including only the $p\Lambda$-final-state interaction as modification of the phase-space results in a strong enhancement at small invariant $p\Lambda$-masses with a vertical profile in the Dalitz plot (Fig.\,\ref{mod_sib} lower right). The data can only be described in a satisfactory way if both contributions are added at amplitude level (Fig.\,\ref{mod_sib} upper right).
\par
\begin{figure}[ht!]
  \hspace{0.7cm}
  \vspace{-0.2cm}
  \includegraphics[width=0.92\textwidth]{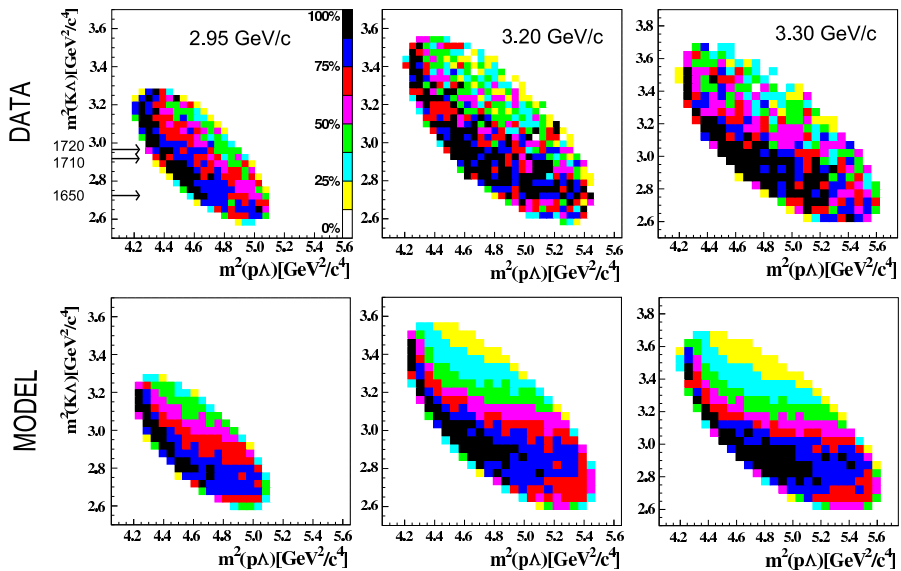}
  \caption {\label{alldata}Dalitz plots of the reaction $pp \rightarrow K^+ \Lambda p$ at $p_{beam}=2.95, 3.20$ and $3.30$\,GeV/c; data (upper) compared to model-fits (lower).}
\end{figure}
\begin{figure}[ht!]
  \includegraphics[scale=0.5]{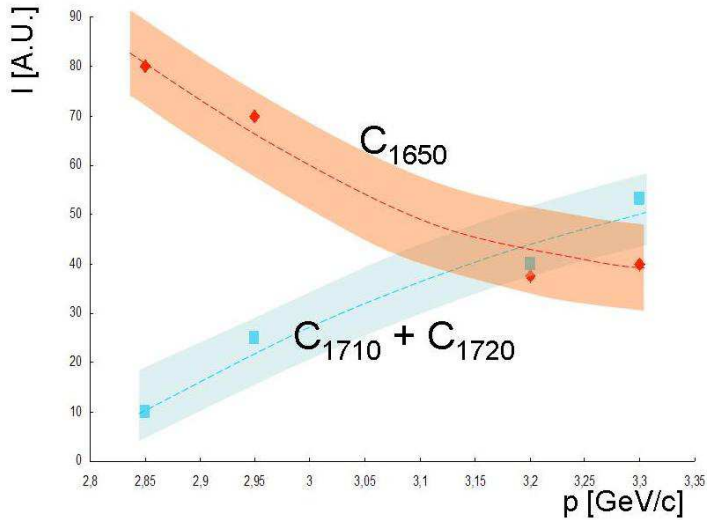}
  \caption {\label{n-star-resonances} Contribution of $N$(1650) compared to the sum of $N(1710)+N(1720)$ as a function of the beam momentum. For comparison the values at $p_{\mathit{beam}}\,=\,2.85$\,GeV/c are added.\cite{COSY-TOF06}}
\end{figure}

In Fig.\ref{alldata} (upper part) Dalitz plots for the experimental data at the three beam momenta of $2.95\,$GeV/c, $3.20\,$GeV/c and 
$3.30\,$GeV/c are shown. To obtain more insight into the various contributions, the data of each beam momentum were compared with the model 
parametrization (see Eq. 1).\\
The strengths of the various contributions were adjusted individually to achieve a best fit for the various Dalitz plots. The results are shown 
in Fig. \ref{alldata} (lower part) and table 2. The data are rather well described by the model fits. The obtained reduced $\chi^2$-values are between 1.4 and 2.0. 
The strength of the contribution of the $N$(1650)-resonance compared to the sum of the contributions of the $N$(1710)- and $N$(1720)-resonances changes strongly with the beam momentum, which is qualitatively expected by the opening of phase-space with excess energy. This is shown in Fig.\,\ref{n-star-resonances}. The dashed curve is obtained by a second order polynomial fit to the extracted strengths of the resonances. The width of the band corresponds to the $3\sigma$ error of the Dalitz plot fit as given in table 2. The amplitude of the non-resonant contribution is smaller by an order of magnitude compared to the sum of the three contributing resonances. This behaviour is not affected by the finite detector resolution, which is more than an order of magnitude better than the resonance width. The influence of the $p\Lambda$-final state interaction is significant even for the highest momentum; the corresponding amplitude $C_{FSI}$ is independent of the beam momentum within the uncertainties (table 2).
\par
From these results it has to be concluded that there is a dominant exchange of non-strange mesons, since only these are able to contribute to the observed leading mechanism via $N^*$-resonances.
In addition, the parameters extracted from the measurement at 2.85\,GeV/c \cite{COSY-TOF06} are included in table 2 and Fig.\,\ref{n-star-resonances}. They fit well in the trend of the actual data.
\par
First indications for contributions of $N^*$-resonances were already discussed for the interpretation of the exclusive bubble chamber data at higher beam energies \cite{Chinowsky68} and for the data of inclusive kaon production experiments \cite{Hogan68} to explain the deviations from phase-space. For the $p\Lambda$-final-state interaction there were first approaches in earlier work to describe the 
deviations from phase-space in inclusive kaon production \cite{Melissinos65}. In comparison to those early
hints, the analysis of the data presented in this paper allows a more precise and unambiguous determination of the 
contribution of the $N^*$-resonances. A quantitative investigation of the $p\Lambda$-final-state interaction in the observed reaction was also 
performed by the COSY-11 experiment a few MeV above the reaction threshold \cite{COSY-1198,COSY-1199}. However, due to 
the small phase-space, data very close to threshold do not allow the clear extraction of contributions of 
$N^*$-resonances even if they contribute strongly. 
\par
In connection with the results reported in this paper concerning the contribution of the different $N^*$-resonances, the model calculations of Shyam \cite{Shyam99} are of special interest. The model, especially used to describe total cross-section data of the reaction $pp \rightarrow K^+ \Lambda p$, is based on the exchange of various non strange mesons and the excitations of the resonances $N$(1650, 1710 and 1720). Similar to the results reported here, a dominant contribution of the $N$(1650) was found close to the threshold. At a beam momentum of about 3.0 GeV/c, the $N$(1710) becomes dominant. This behaviour agrees with the results shown in Fig.\,\ref{n-star-resonances}. Further insight to the role of $N^*$ resonances could be expected by performing a partial wave analysis of these data in a similar way as it was shown for $\gamma$ induced reactions \cite{Sarantsev2005}.
\par
\begin{figure}[ht!]
  \includegraphics[width=\textwidth]{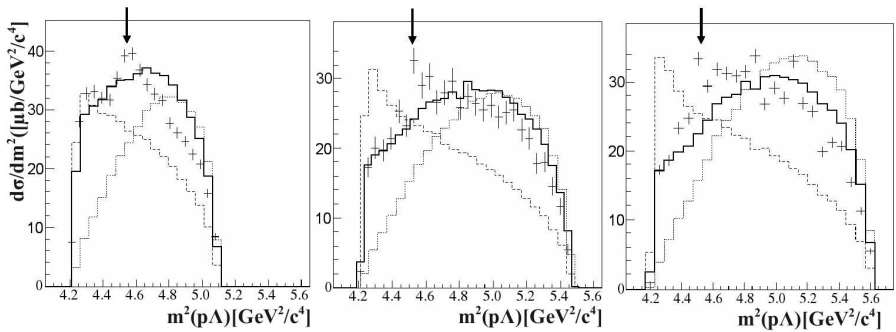}
  \caption{\label{subsysteme} \footnotesize {Projections to the $p\Lambda$-subsystem at $p_{\mathit{beam}}\,=$\,2.95, 3.20 and 3.30\,GeV/c (left to right) together with results of the model of Sibirtsev (solid line). The contributions from the individual components are shown: only resonances (dotted line) and only $p\Lambda$-final-state interaction (dashed line). The arrows indicate the $p\Sigma^{0}$-threshold.}} 
  \end{figure}

The investigation of the Dalitz plots and the projections on the masses of the two-body systems gives in addition access to study the effect of the onset of $\Sigma$ production on the detected $K^+\Lambda p$ final state.
In a semi-inclusive K-production experiment at SACLAY \cite{Sie94} evidence for a cusp-effect in the $\Lambda$-cross-section at the $p\Sigma^{0}$-threshold was reported. An indication of a cusp in the $p\Lambda$-spectrum was already reported in \cite{COSY-TOF06}. Further evidence is now given from the measurements at higher beam momenta, as shown in Fig.\,\ref{subsysteme}, where in the projections to the p$\Lambda$ system an enhancement at the p$\Sigma^0$ threshold is visible in all three spectra.

As expected from the Dalitz plot itself, the comparison of the projections,
plotted in Fig.\,\ref{subsysteme} also shows a reasonable agreement between the 
data and the adjusted model, apart from the region of the $p\Sigma^0$ threshold. None of the individual components alone is able to describe the behaviour of the data in both projections. 
\section{Summary}
The reaction $\mathit{pp}\,\rightarrow\,K^+{\!\Lambda}p$ was
measured exclusively at the three beam momenta of $2.95$, $3.20$ 
and $3.30$\,GeV/c. From the Dalitz plot analysis of the data at three excess energies,
detailed information on the reaction mechanism could be extracted. The analysis
shows evidence for a dominant contribution of $N^*$-resonances. From this it has to be concluded that the exchange of
non-strange mesons is the leading process in the energy range near threshold. The strength of the contribution of the $N$(1650)-resonance compared to the sum of the contributions of the $N$(1710)- and $N$(1720)-resonances changes decisively with the beam momentum. In addition, a strong influence of the $p\Lambda$-final-state interaction on the Dalitz plot distribution was found even at about $300$\,MeV above threshold.

\section{Acknowledgements}
We would like to thank very much the COSY accelerator team for the preparation of the excellent 
proton beam and the good cooperation during the beam time.
We are very grateful to A. Sibirtsev for many fruitful discussions and the
preparation of his model for Dalitz plot analyses.
This work is based in part on the doctoral thesis of Wolfgang Schroeder.
We gratefully acknowledge support from
the German BMBF and the FZ J\"ulich.

\end{document}